\documentclass[twocolumn]{aastex62}
\usepackage{multirow}
\usepackage{float}
\usepackage[applemac]{inputenc}
\usepackage{natbib}
\usepackage[]{threeparttable}
\newcommand{\hi}{H\,{\sc i}}

\shortauthors{\textsc{Mutlu-Pakd{\rlap{\.}\i}l} et al.}

\begin{document}

\title{{\it Hubble Space Telescope} Observations of NGC~253 Dwarf Satellites: Discovery of Three Ultra-faint Dwarf Galaxies \footnote{This paper includes data gathered with the 6.5~m Magellan Telescope at Las Campanas Observatory, Chile.}}

\correspondingauthor{B. Mutlu-Pakdil}
\email{burcinmp@uchicago.edu}

\author[0000-0001-9649-4815]{\textsc{Bur\c{c}{\rlap{\.}\i}n Mutlu-Pakd{\rlap{\.}\i}l}}
\affil{Kavli Institute for Cosmological Physics, University of Chicago, Chicago, IL 60637, USA}
\affil{Department of Astronomy and Astrophysics, University of Chicago, Chicago IL 60637, USA}

\author[0000-0003-4102-380X]{David J. Sand}
\affil{Steward Observatory, University of Arizona, 933 North Cherry Avenue, Tucson, AZ 85721, USA}

\author[0000-0002-1763-4128]{Denija Crnojevi\'{c}}
\affil{University of Tampa, 401 West Kennedy Boulevard, Tampa, FL 33606, USA}

\author[0000-0002-5434-4904]{Michael G. Jones}
\affil{Steward Observatory, University of Arizona, 933 North Cherry Avenue, Tucson, AZ 85721, USA}

\author[0000-0003-2352-3202]{Nelson Caldwell}
\affil{Center for Astrophysics, Harvard \& Smithsonian, 60 Garden Street, Cambridge, MA 02138, USA}

\author[0000-0001-8867-4234]{Puragra Guhathakurta}
\affiliation{University of California Observatories/Lick Observatory, University of California, Santa Cruz, CA 95064, USA}

\author[0000-0003-0248-5470]{Anil C. Seth}
\affil{University of Utah, 115 South 1400 East Salt Lake City, UT 84112-0830, USA}

\author[0000-0002-4733-4994]{Joshua D. Simon}
\affiliation{Observatories of the Carnegie Institution for Science, 813 Santa Barbara Street, Pasadena, CA 91101, USA}

\author[0000-0002-0956-7949]{Kristine Spekkens}
\affil{Department of Physics and Space Science, Royal Military College of Canada P.O. Box 17000, Station Forces Kingston, ON K7K 7B4, Canada}
\affil{Department of Physics, Engineering Physics and Astronomy, Queen's University, Kingston, ON K7L 3N6, Canada}

\author[0000-0002-1468-9668]{Jay Strader}
\affil{Department of Physics and Astronomy, Michigan State University,East Lansing, MI 48824, USA}

\author[0000-0001-6443-5570]{Elisa Toloba}
\affil{Department of Physics, University of the Pacific, 3601 Pacific Avenue, Stockton, CA 95211, USA}

\begin{abstract}

We present deep {\it Hubble Space Telescope} imaging of five faint dwarf galaxies associated with the nearby spiral NGC~253 (D$\approx$3.5 Mpc).  Three of these are newly discovered ultra-faint dwarf galaxies, while all five were found in the Panoramic Imaging Survey of Centaurus and Sculptor (PISCeS), a Magellan$+$Megacam survey to identify faint dwarfs and other substructures in resolved stellar light around massive galaxies outside of the Local Group.  Our {\it HST} data reach $\gtrsim$3~magnitudes below the tip of the red giant branch for each dwarf, allowing us to derive their distances, structural parameters, and luminosities.  All five systems contain predominantly old, metal-poor stellar populations (age$\sim$12~Gyr, [M/H]$\lesssim$$-$1.5) and have sizes ($r_{h}$$\sim$110--3000~pc) and luminosities ($M_V$$\sim$$-7$ to $-12$~mag) largely consistent with Local Group dwarfs. The three new NGC~253 satellites are among the faintest systems discovered beyond the Local Group. We also use archival \hi~data to place limits on the gas content of our discoveries. Deep imaging surveys such as our program around NGC~253 promise to elucidate the faint end of the satellite luminosity function and its scatter across a range of galaxy masses, morphologies, and environments in the decade to come. 

\end{abstract}


\section{Introduction} \label{sec:intro}

Low mass dwarf galaxies are an important probe at the intersection of the smallest dark matter halos and the astrophysical processes that shape galaxy formation.  In the $\Lambda$+Cold Dark Matter ($\Lambda$CDM) model for structure formation, galaxies grow hierarchically within DM halos, but quantitatively verifying this picture on dwarf galaxy scales has proved challenging \citep{Bullock17}.  These challenges include the `missing satellites problem' \citep[e.g.][]{Moore99,Klypin99}, `too big to fail' \citep[e.g.][]{BK11,BK12}, and the apparent planes of satellites around nearby galaxies \citep[e.g.][]{Pawlowski12,Ibata13,Muller18}. 

Significant progress has been made in addressing these small-scale $\Lambda$CDM challenges on the theoretical front, as the inclusion of realistic baryonic physics into simulations of Milky Way-like galaxies can broadly reproduce the properties of the dwarf galaxies in the Local Group \citep[e.g.,][]{Brooks13,Sawala16,Wetzel16,Samuel20,Engler21}.  At the same time, the number and diversity of observed satellites around the Milky Way (MW) continues to grow \citep[most recently][]{Mau20,Cerny20}.  The MW will remain an essential proving ground for understanding the astrophysics and cosmological implications of the very faintest dwarf galaxy satellites because of the detail and depth with which it can be studied (see \citealt{Simon2019} for a recent review).

Despite the progress in the Local Group, its detailed study  will not be sufficient for verifying the $\Lambda$CDM model on small scales, and there is a danger of `overtuning' the models to match local observations alone.  Fortunately, in the coming decade we will greatly expand our understanding of faint satellites not just within our own Local Group, but well into the Local Volume to sample primary halos with a range of masses, morphologies and environments all the way down to the ultra-faint dwarf galaxy scale \citep[e.g. see the recent simulations and discussion in ][]{MutluPakdil21}.  Indeed, the census of faint dwarfs around nearby galaxies is well underway using resolved and unresolved imaging \citep[e.g.,][]{Chiboucas13,Sand14,Sand15b,Crnojevic14,Crnojevic16,Crnojevic19,Carlin16,Carlin2020,Toloba16,Danieli17,Smercina18,Bennet17,Bennet19,Bennet20,Carlstenyada,Davis21,Drlica-Wagner2021,Carlsten21,Garling21}, as well as spectroscopic surveys around MW analogs at larger distances \citep{Geha17,Mao20}.  These data are already yielding new challenges, with simulations of MW-like galaxies showing a higher fraction and number of quiescent satellites with respect to the observations \citep{Karunakaran21}.

In this work, we present {\it Hubble Space Telescope} ({\it HST}) observations of dwarf galaxy candidates around NGC~253 ($D$$\approx$$3.5$~Mpc; \citealt{Radburn-Smith2011}, a total stellar mass of $\approx$$4.4$$\times$$10^{10}$~$M_{\odot}$; \citealt{Bailin2011}), the principal galaxy of the nearby Sculptor group.  These dwarf satellites were discovered as part of the Panoramic Imaging Survey of Centaurus and Sculptor  \citep[PISCeS;][]{Sand14,Toloba16,Crnojevic14,Crnojevic16,Crnojevic19,Hughes21,MutluPakdil21}, and include three new discoveries of ultra-faint dwarf satellites (adopting the definition of $M_V$$\gtrsim$$-$7.7 in \citealt{Simon2019}) around NGC~253.  In Section~\ref{sec:survey} we give an overview of PISCeS and the search for satellites around NGC~253, and in Section~\ref{sec:observations} we present the {\it HST} observations of our new NGC~253 dwarf galaxy candidates.  Next, in Section~\ref{sec:properties} we measure the properties of our dwarf sample, including their stellar population, distance, gas content and structural parameters.  We discuss and conclude in Section~\ref{sec:conclude}.

\section{PISCES and Discovery of Three Dwarf Satellites Around NGC~253} \label{sec:survey}

\begin{figure*}
\centering
\fbox{\includegraphics[width = 5.4in]{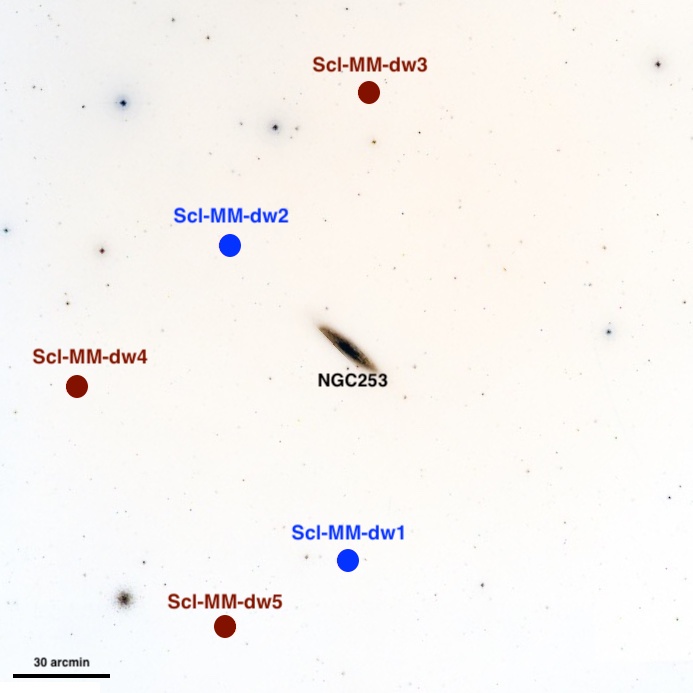}}\\
\begin{picture}(97,97)
\put(0,-4){\includegraphics[width = 1.4in]{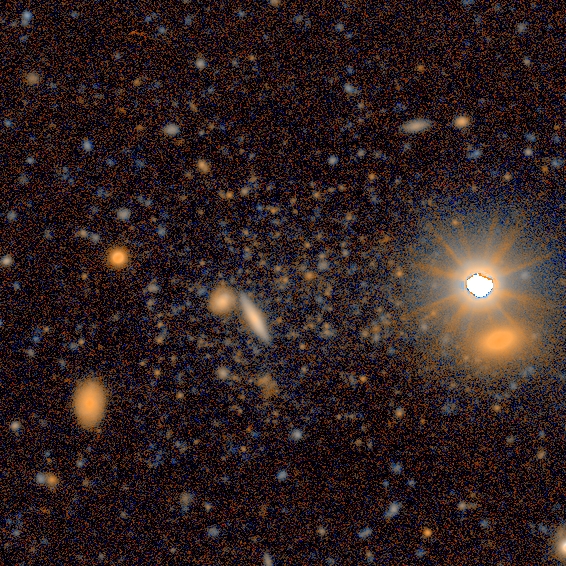}}
\put(25,4){\color{white} \textbf{Scl-MM-dw1}}
\put(5,85){\color{orange} \textbf{Magellan/Megacam}}
\end{picture}
\begin{picture}(97,97)
\put(0,-4){\includegraphics[width = 1.4in]{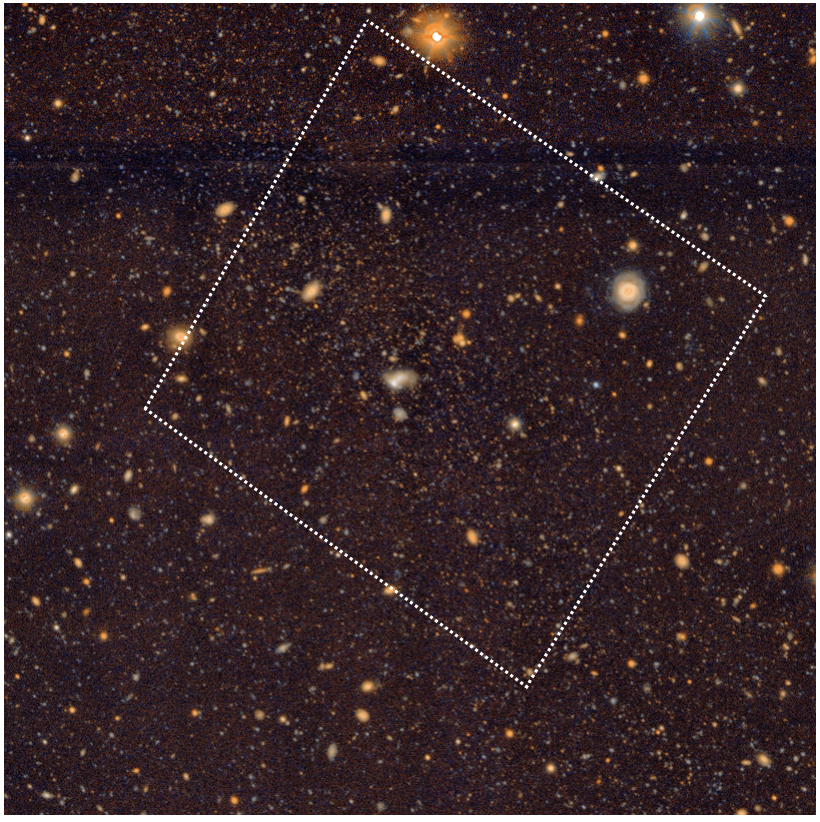}}
\put(25,4){\color{white} \textbf{Scl-MM-dw2}}
\put(5,85){\color{orange} \textbf{Magellan/Megacam}}
\end{picture}
\begin{picture}(97,97)
\put(0,-4){\includegraphics[width = 1.4in]{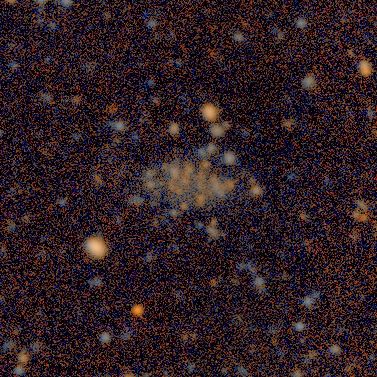}}
\put(25,4){\color{white} \textbf{Scl-MM-dw3}}
\put(5,85){\color{orange} \textbf{Magellan/Megacam}}
\end{picture}
\begin{picture}(97,97)
\put(0,-4){\includegraphics[width = 1.4in]{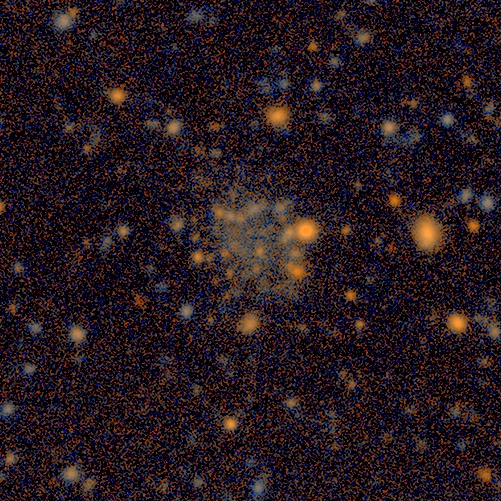}}
\put(25,4){\color{white} \textbf{Scl-MM-dw4}}
\put(5,85){\color{orange} \textbf{Magellan/Megacam}}
\end{picture}
\begin{picture}(97,97)
\put(0,-4){\includegraphics[width = 1.4in]{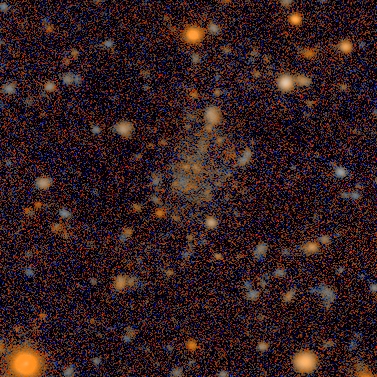}}
\put(25,4){\color{white} \textbf{Scl-MM-dw5}}
\put(5,85){\color{orange} \textbf{Magellan/Megacam}}
\end{picture}
\\
\begin{picture}(97,97)
\put(2,-8){\includegraphics[width = 1.4in]{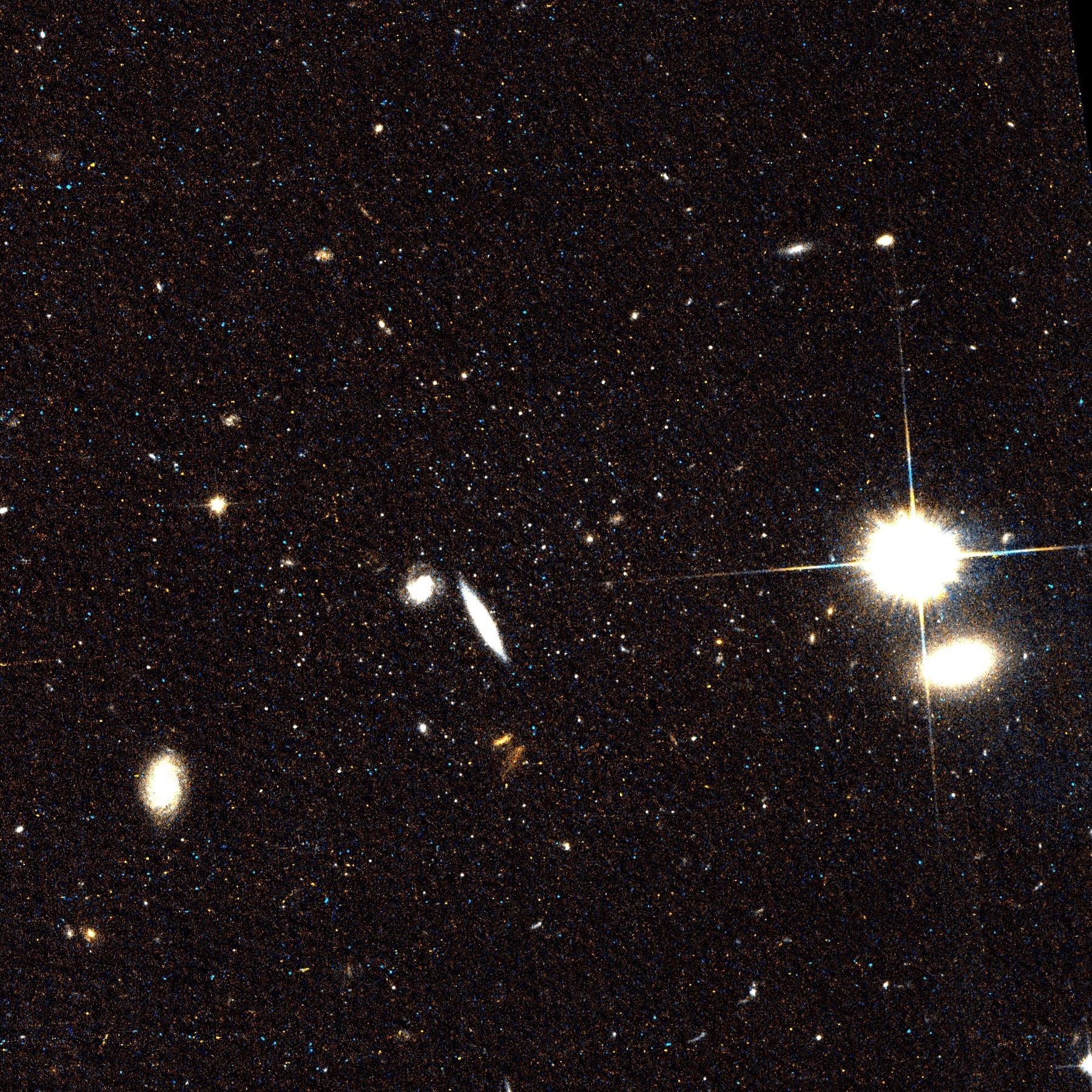}}
\put(25,0){\color{white} \textbf{Scl-MM-dw1}}
\put(50,80){\color{orange} \textbf{HST/ACS}}
\end{picture}
\begin{picture}(97,97)
\put(2,-8){\includegraphics[width = 1.4in]{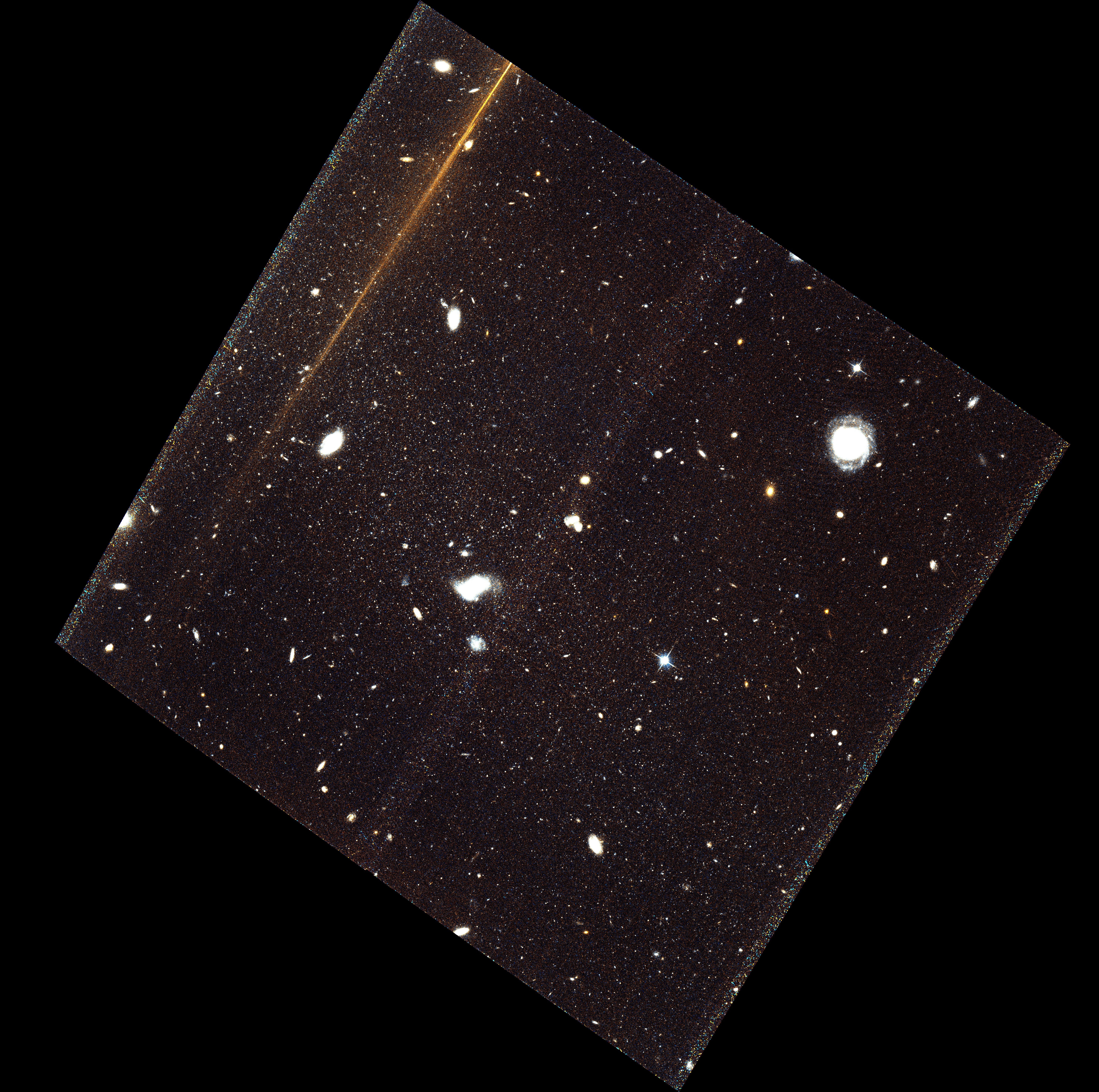}}
\put(25,0){\color{white} \textbf{Scl-MM-dw2}}
\put(50,80){\color{orange} \textbf{HST/ACS}}
\end{picture}
\begin{picture}(97,97)
\put(2,-8){\includegraphics[width = 1.4in]{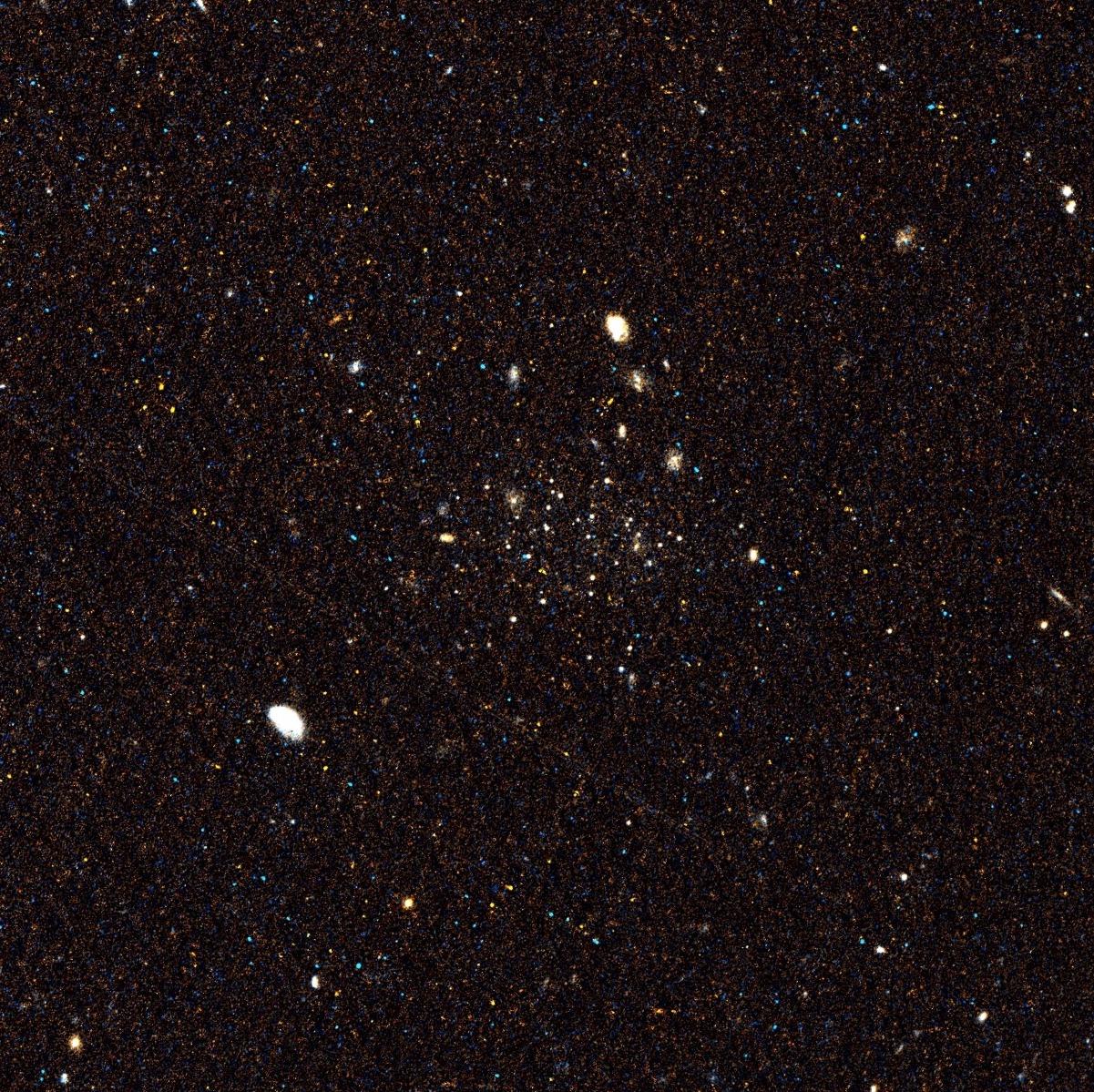}}
\put(25,0){\color{white} \textbf{Scl-MM-dw3}}
\put(50,80){\color{orange} \textbf{HST/ACS}}
\end{picture}
\begin{picture}(97,97)
\put(2,-8){\includegraphics[width = 1.4in]{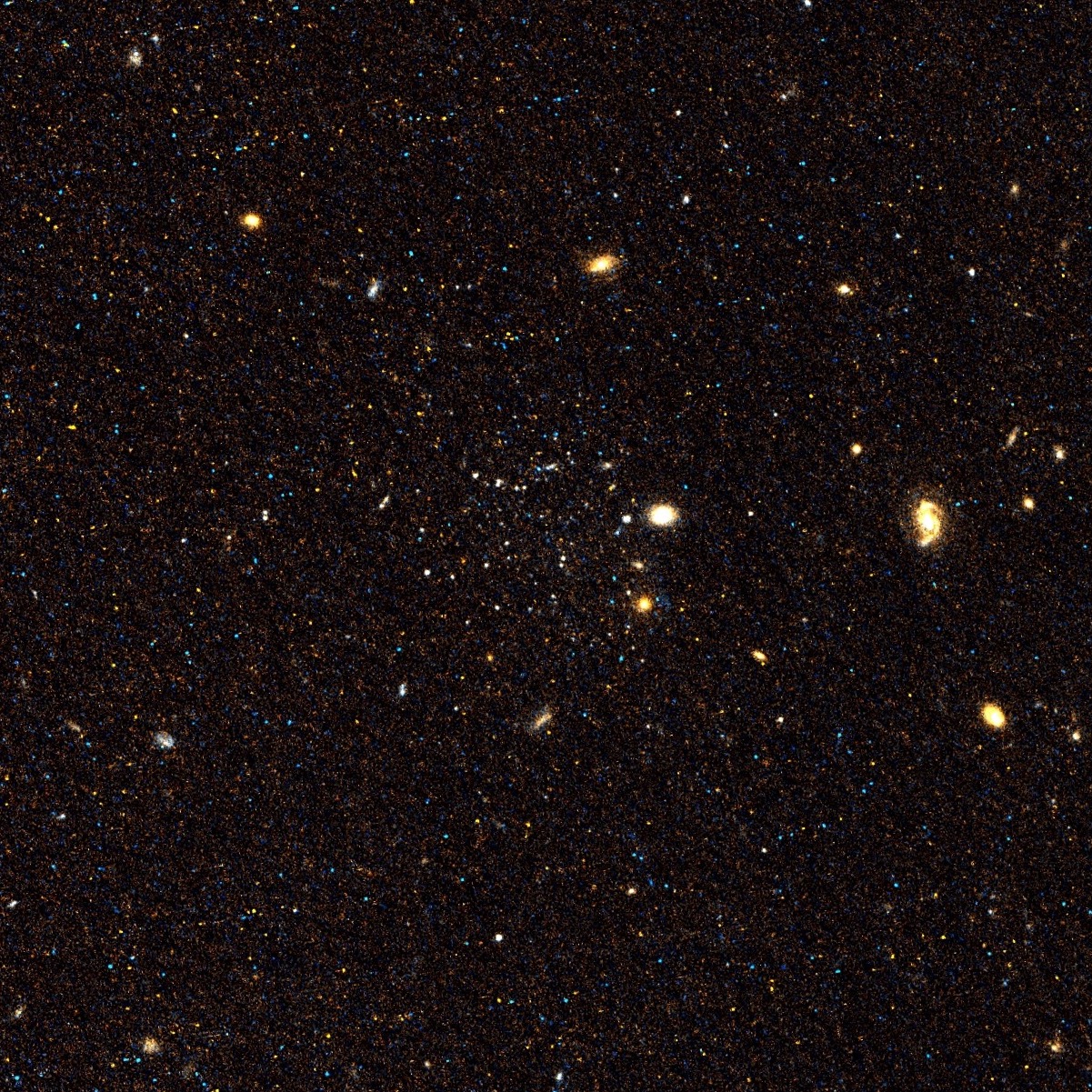}}
\put(25,0){\color{white} \textbf{Scl-MM-dw4}}
\put(50,80){\color{orange} \textbf{HST/ACS}}
\end{picture}
\begin{picture}(97,97)
\put(2,-8){\includegraphics[width = 1.4in]{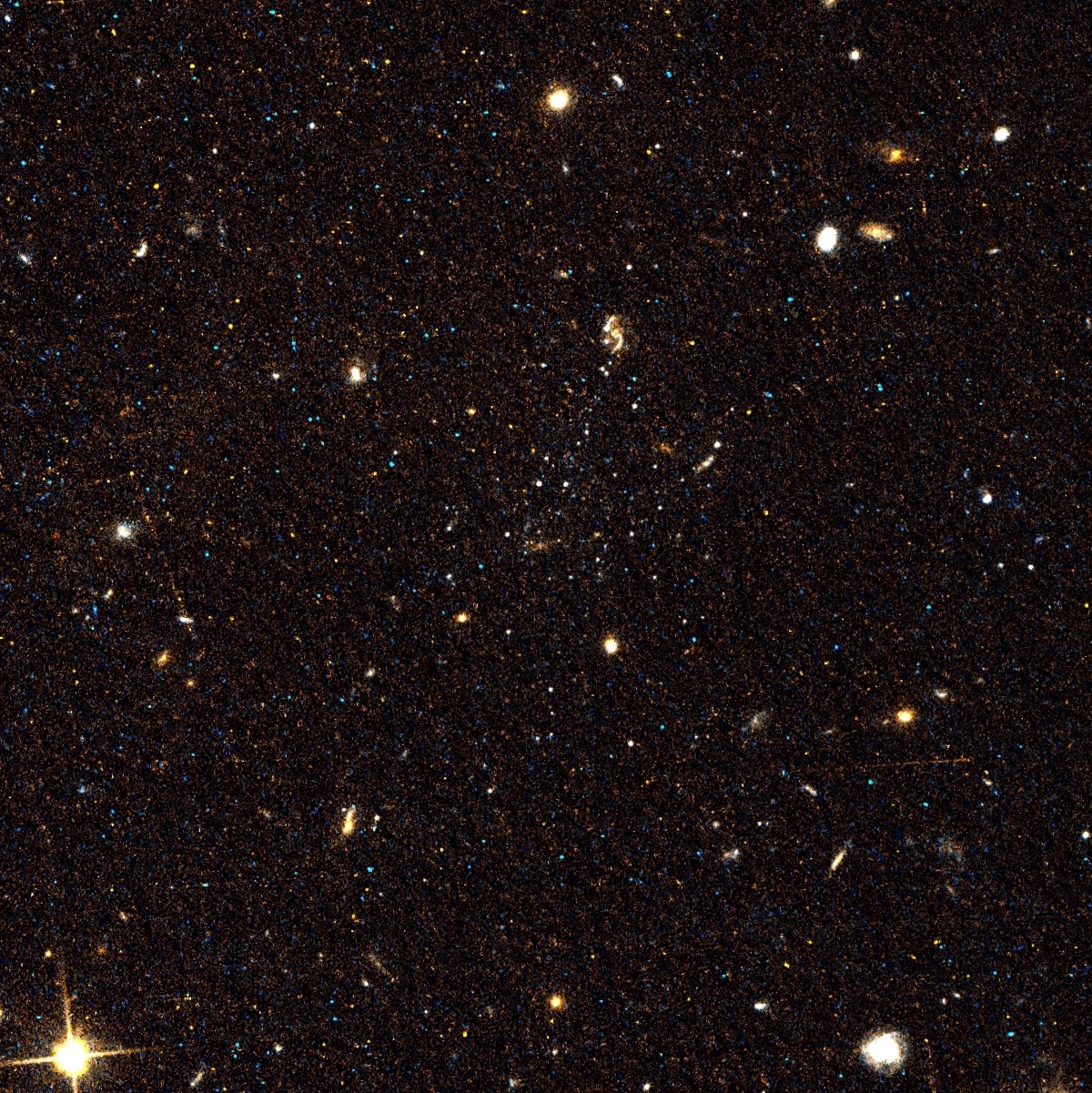}}
\put(25,0){\color{white} \textbf{Scl-MM-dw5}}
\put(50,80){\color{orange} \textbf{HST/ACS}}
\end{picture}
\vspace{0.075in}
\caption{Top panel: DSS image centered on NGC~253, showing the area explored by PISCeS, extending up to 100~kpc from the center. Solid blue circles represent the position of dwarfs previously discovered in PISCeS (Scl-MM-dw1, \citealt{Sand14}; Scl-MM-dw2, \citealt{Toloba16}) while red circles represent new dwarf galaxies reported in this study. North is up, and East to the left. Middle (Bottom) panel: RGB false color Magellan/Megacam (HST/ACS) images of our dwarf galaxy discoveries.
The image cutout sizes of Scl-MM-dw1 are $1.5\arcmin\times1.5\arcmin$. The 
Megacam cutout size of Scl-MM-dw2 is $6\arcmin\times6\arcmin$. The ACS FoV, shown with the white dotted box, is too small to cover the entire body of Scl-MM-dw2 ($r_h=3.2\pm0.6$~arcmin), therefore we show the entire ACS FoV for Scl-MM-dw2 in the bottom panel. The size of the other cutouts is $1\arcmin\times1\arcmin$. Note that 1~arcmin corresponds to 1.02~kpc at 3.5~Mpc.
\label{fig:position}}
\end{figure*}

The Panoramic Imaging Survey of Centaurus and Sculptor (PISCeS) is a Magellan$+$Megacam survey  to search for faint satellites and signs of hierarchical structure formation in the halos of two nearby galaxies of different morphologies in two environments substantially different from the Local Group -- the starbursting spiral NGC~253 in a loose group of galaxies ($D\approx3.5$~Mpc; \citealt{Radburn-Smith2011}) and the elliptical NGC~5128, or Centaurus~A (Cen~A) in a relatively rich group ($D\approx3.8$~Mpc; \citealt{Harris2010}). The PISCeS campaign has led to the discovery of 11 new satellite candidates around Cen~A and several previously unknown streams and shells \citep{Crnojevic16}, which were later followed up with {\it HST} data and confirmed \citep{Crnojevic19}. 

As part of PISCeS, we have observed 82 Megacam fields around NGC~253, which reach out to a projected radius of $\sim$$100$~kpc ($\sim$$1/3$ of its virial radius; see Figure~\ref{fig:position} for the survey footprint). Megacam has a $\sim24\arcmin\times24\arcmin$ field of view (FoV) and a binned pixel scale of 0.16\arcsec. PISCeS typically observes each field for $6\times300$~s in each of the $g$ and $r$ bands to achieve image depths of $g,r\approx26.5$~mag, which is $\sim$2 magnitudes below the tip of the red giant branch (TRGB) at the distance of NGC~253. The median seeing throughout the survey has been $\sim$$0.8\arcsec$ in both bands. The data are reduced in a standard way by the Smithsonian Astrophysical Observatory Telescope Data Center (see \citealt{McLeod15,Crnojevic16}, for further details).

We visually inspect all the images, searching for spatially compact overdensities of stars which have some signs of diffuse light as well. In the early stages of the survey, two faint NGC~253 satellites were reported in \citet[][Scl-MM-dw1]{Sand14} and  \citet[][Scl-MM-dw2\footnote{This galaxy was independently discovered by \citet{Romanowsky2016}, and was named NGC~253-dw2 in their work.}]{Toloba16}. Here we report the discoveries of three new faint, diffuse and elongated satellite galaxies, which we dub Scl-MM-dw3\footnote{During the preparation of this paper, we learned of an independent discovery of the same object using the Dark Energy Survey data \citep[DES,][]{Martinez-Delgado21}. The authors named it Donatiello~II.}, Scl-MM-dw4 and Scl-MM-dw5, in accordance with our prior work in the Sculptor group. The locations and properties of these dwarfs are given in Figure~\ref{fig:position} and Table~\ref{tab:dwarfs}.  In the current work, we focus on the {\it HST} color-magnitude diagrams and derived properties of this set of five faint dwarf galaxies. 
These objects appear to represent a complete sample of the dwarf galaxies detectable in the PISCeS data set; the overall detection efficiency and a satellite luminosity function will be presented in a forthcoming work.

\vspace{0.075in}
\section{HST Observations and Photometry} \label{sec:observations}

Including the discoveries of Scl-MM-dw1 and Scl-MM-dw2, PISCeS has uncovered 5 dwarf satellites around NGC~253 in total (see Table~\ref{tab:dwarfs}).  We obtained {\it HST} follow-up observations of these dwarfs with the Wide Field Channel (WFC) of the Advanced Camera for Surveys (ACS). Most of the targets were observed as part of the program GO-15938 (PI: Mutlu-Pakdil), with the exception of Scl-MM-dw2, which was observed as part of program GO-14259 (PI: Crnojevi\'{c}). Each target was observed for a total of one orbit (two orbits for Scl-MM-dw2) in the F606W and F814W filters, yielding exposure times of $\sim$1100 and 2500~s per filter, for one and two orbits, respectively.

We performed point-spread function photometry on the pipeline-produced FLC images with the latest version (2.0) of DOLPHOT \citep{Dolphin2000}, largely using the recommended prescriptions. The initial photometry is  culled with the following criteria: the sum of the crowding parameters in the two bands is $<$$1$, the squared sum of the sharpness parameters in the two bands is $<$$0.075$, and the signal-to-noise ratio is $>$$4$ and object-type is $\leq$$2$ in each band. We corrected for Milky Way extinction on a star-by-star basis using the \citet{Schlegel98} reddening maps with the coefficients from \citet{Schlafly11}. The extinction-corrected photometry is used throughout this work. We derived completeness and photometric uncertainties using $\sim$100,000 artificial star tests per field, measured with the same photometric routines used to create the photometric catalogs. Our one-orbit {\it HST} data are 50\% (90\%) complete at F606W$\sim$27.1~(26.4)~mag and F814W$\sim$26.4~(25.7)~mag, while the two-orbit data set of Scl-MM-dw2 reaches 50\% (90\%) completeness at F606W$=$27.9 (27.2)~mag and F814W$=$27.1 (26.6)~mag.

\section{Properties of NGC~253 Dwarfs} \label{sec:properties}

\subsection{Color-Magnitude Diagram \label{sec:cmd}}

\begin{figure*}
\centering
\includegraphics[width=\linewidth]{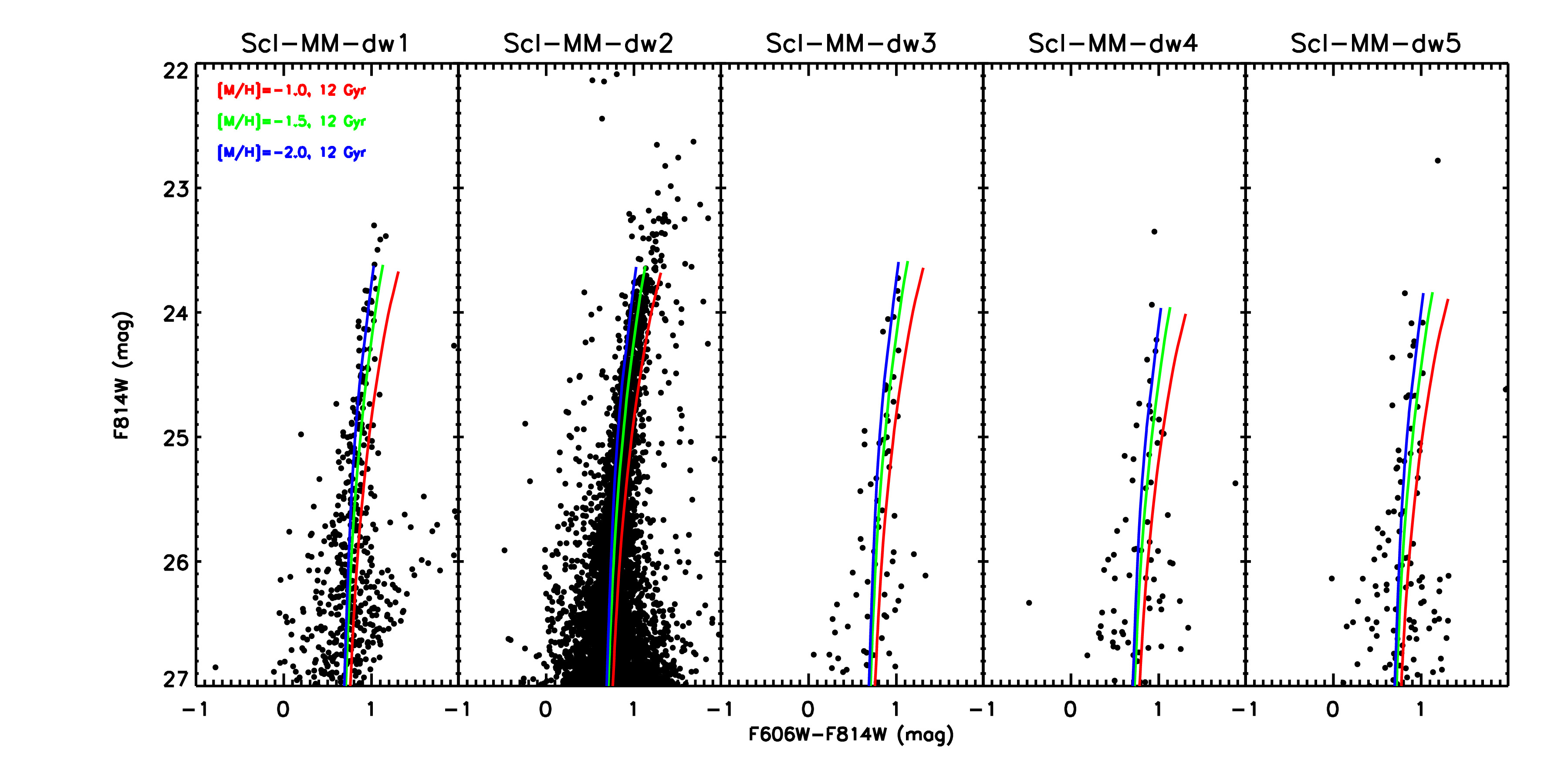}
\vspace{-0.35in}
\caption{{\it HST} CMDs showing the stars within $2\times r_{h}$ of each dwarf galaxy except Scl-MM-dw2 (its CMD includes all stars within the ACS FoV, and does not cover the entire body of Scl-MM-dw2, $r_h=3.2\pm0.6$~arcmin). The blue, cyan, and red lines indicate the PARSEC isochrones for 12 Gyr and [M/H]$=-2.0$~dex, $-1.5$~dex, and $-1.0$~dex, respectively. We shift each isochrone by the best-fit distance modulus that we derive in Section~\ref{sec:dist}. \label{fig:cmd}}
\end{figure*}

Figure~\ref{fig:cmd} shows the {\it HST} color-magnitude diagrams (CMDs) of the five dwarfs, which include stars within two half-light radii. Note that the ACS FoV is too small to cover the entire body of Scl-MM-dw2 ($r_h=3.2\pm0.6$~arcmin), therefore we include all stars in the entire ACS FoV for Scl-MM-dw2. Overplotted as blue, cyan, and red lines are the PARSEC isochrones \citep{Bressan2012} for 12~Gyr and [M/H]$=-2.0$~dex, $-1.5$~dex, and $-1.0$~dex, respectively. Each dwarf is clearly resolved into its constituent red giant branch (RGB) stars in the {\it HST} data, and shows old, metal-poor stellar populations at roughly the distance to NGC~253 (see Section~\ref{sec:dist}). 

The three new dwarf galaxies we report here are extremely faint, and have sparsely populated RGBs. They are consistent with old, metal-poor systems comprised of predominantly ancient stellar populations, similar to the ultra-faint dwarfs in the MW and M31 \citep[e.g.,][]{Brown2012,Brown2014,Martin2016,Simon2019}.

The old RGB in Scl-MM-dw2 closely follows the isochrone of [M/H]$=-1.5$ (green line), which is what we would expect for its luminosity ($M_V=-12.1$) based on the luminosity-metallicity relationship \citep{Kirby2013}. This implies that the stripping of stars in the satellite suggested by \citet{Toloba16} should be moderate. However, we note that the relationship has a lot of scatter; and given its large size and elongation, the object is likely undergoing a tidal interaction (see Section~\ref{sec:conclude}).

\begin{figure*}
\centering
\includegraphics[width=\linewidth]{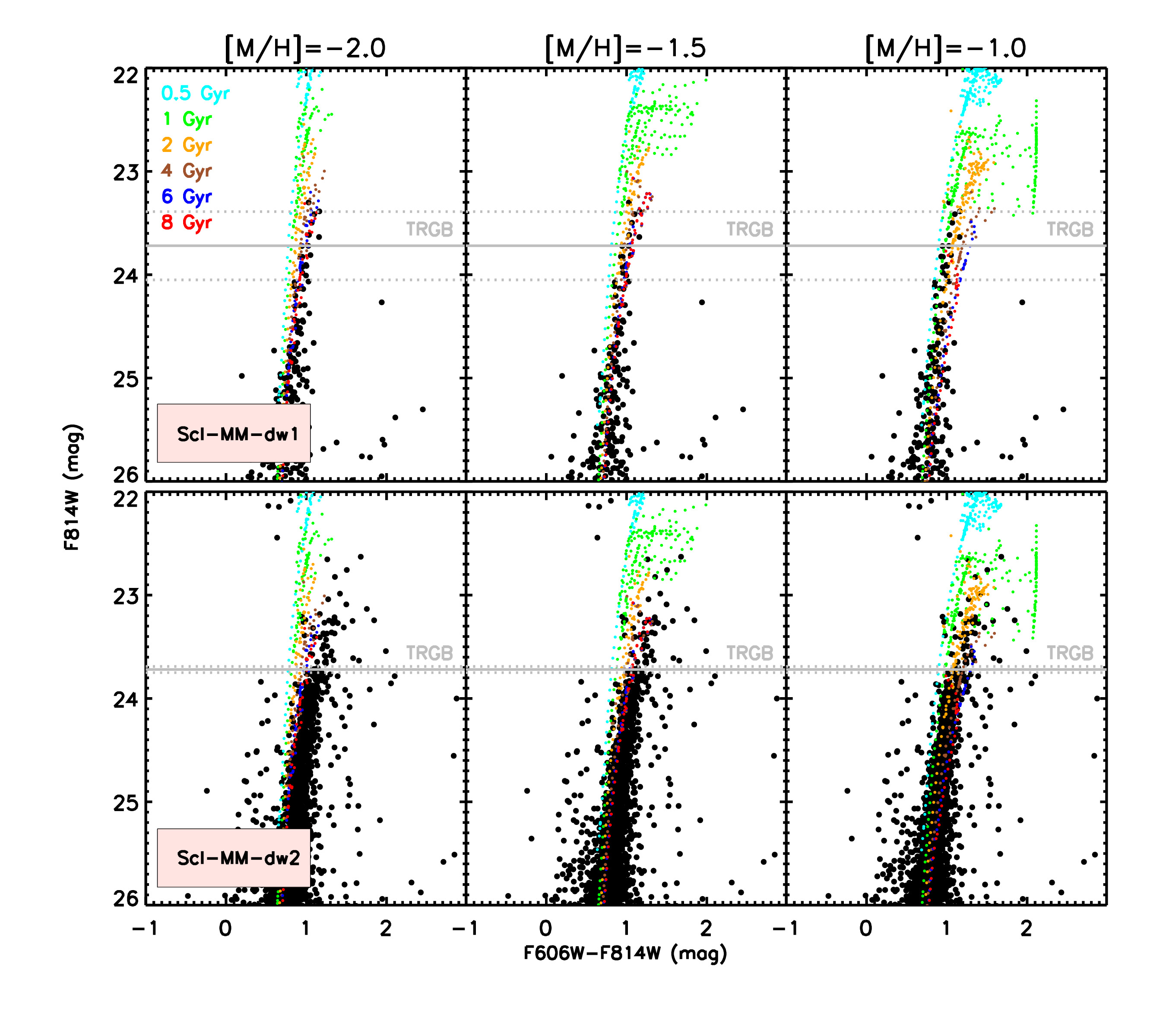}
\vspace{-0.5in}
\caption{To better understand the AGB stars in Scl-MM-dw1 (top panel) and Scl-MM-dw2 (bottom panel), we plot the AGB phase of PARSEC isochrones for the indicated ages over their stars (black points) for different metallicities (left: [M/H]$=-2$, middle: [M/H]$=-1.5$, right: [M/H]$=-1.0$). 
The gray horizontal line indicates the best fit for the TRGB, and the dotted lines represent the $1\sigma$ uncertainty. We roughly look for a match between the isochrones and stars that are above the TRGB. The match suggests a population of $\sim$1--2~Gyr and [M/H]$\sim-1.0$~dex for the AGB stars in Scl-MM-dw2, and a population of $\gtrsim6$--$8$~Gyr for the AGB stars in Scl-MM-dw1.  \label{fig:cmd2}}
\end{figure*}

In the ground-based Magellan imaging \citep{Sand14,Toloba16}, both Scl-MM-dw1 and Scl-MM-dw2 showed some evidence of a younger asymptotic giant branch (AGB) stellar population, which is confirmed by our deeper {\it HST} follow-up imaging. Figure~\ref{fig:cmd2} shows the AGB phase for isochrones with a range of ages and metallicities for both of these galaxies. Scl-MM-dw1 seems to contain only a handful of such luminous, possibly intermediate--age populations; these stars complicate its TRGB identification, resulting in a larger uncertainty in the TRGB value (F814W$_{\rm TRGB}$$=$$23.72\pm0.33$~mag, see Section~\ref{sec:dist}). This galaxy does not seem to contain populations younger than $\sim6-8$~Gyr: the low stellar mass of Scl-MM-dw1 and the inherent stochasticity of the AGB phase make it difficult to constrain the amount of possible intermediate--age star formation. In Scl-MM-dw2, the presence of bright AGB stars stretching $\sim$1 magnitude above the TRGB requires young populations. The best fit stellar models suggest these stars are $\sim$1--2~Gyr old, and are somewhat more metal rich than the bulk of stars along the RGB. Based on the luminosity functions from the PARSEC library and the ratio of AGB to RGB stars in our {\it HST} FoV, we find that $\sim$10\% of the stellar mass is in a $\sim$1--2~Gyr population.

\subsection{Distance \label{sec:dist}}
We measure distances to our targets using the TRGB method \citep[e.g.,][]{Lee1993,Salaris2002,Rizzi2007}, which relies on the fixed luminosity of the core helium ignition stage for old stellar populations \citep[e.g.,][]{Serenelli2017}. Details on the TRGB magnitude recovery method can be found in \citet{Crnojevic19}. In short, we compute the observed luminosity function for RGB stars, and then fit it with a model luminosity function, convolved with the appropriate photometric uncertainty and incompleteness function as derived from our artificial star tests. Note that NGC~253 is located at very high galactic latitude ($b=-88^{\circ}$) such that the MW star contamination is very low. 

The TRGB values, the distance moduli, and the distances for our targets are reported in Table~\ref{tab:dwarfs}. For Scl-MM-dw1 and Scl-MM-dw2, the distances derived from the {\it HST} data set are consistent with those derived from the discovery Magellan data set within the uncertainties; the large uncertainty for Scl-MM-dw1 is due to the possible presence of a handful of luminous AGB stars. 

For the remaining three dwarfs, the paucity of RGB stars prevented the code from converging to a reasonable result (as already described in, e.g., \citealt{carlin21}). We thus apply a simple Sobel filter edge-detection algorithm to these dwarfs (following the prescription from \citealt{sakai96}), and computed the related uncertainties with a Monte Carlo (MC) calculation, varying the position of the stars in the CMD within their photometric errors.
However, for our three faint dwarfs, it seems likely that the shot noise in the CMD is a larger effect than the photometric uncertainties. To simulate the shot noise, we first produce a well-populated CMD (of $\sim$20,000 stars) in {\it HST} filters, including our completeness and photometric uncertainties, by using an old, metal-poor isochrone (12~Gyr, [M/H]$=-2.0$~dex) and its associated luminosity function. We then randomly select the observed number of stars from this artificial CMD and measure the typical offset between the brightest simulated star and the location of the TRGB via 1000 realizations. We find that the offset is $0.09$~mag for a faint system like Scl-MM-dw3 or Scl-MM-dw4 ($M_{V}$$\sim$$-7.25$, see Table~\ref{tab:dwarfs} and Section~\ref{sec:lum}), and $0.05$~mag for a Scl-MM-dw5-like dwarf ($M_{V}$$=$$-7.50$), while the offset is consistent with zero for a Scl-MM-dw1-like dwarf ($M_{V}$$=$$-8.75$). This clearly shows that the TRGB technique alone is no longer reliable for ultra-faint dwarfs, and the shot noise should be properly accounted for at these magnitudes. We add these offsets to our bright-end of the MC uncertainties as the measured TRGB will always be below the true TRGB because of the paucity of RGB stars in the faint dwarfs.    
 
The good agreement of TRGB distances with the distance of NGC~253 (e.g., \citealt{Radburn-Smith2011}, who found $m-M =27.70\pm0.07$) firmly establishes their membership with NGC~253.

\subsection{Structural Properties \label{sec:str}}
We derive structural parameters (including half-light radius $r_h$, ellipticity, and position angle) for the dwarfs using the maximum-likelihood (ML) method of \citet{Martin08}, as implemented by \citet{Sand2009}. In our analysis, we select stars consistent with an old, metal-poor isochrone in color-magnitude space after taking into account photometric uncertainties, within our 90\% completeness limit. We inflate the uncertainty to 0.1~mag when the photometric errors are $<0.1$~mag for the purpose of selecting stars to go into our ML analysis. We fit a standard exponential profile plus constant background to the data, and summarize the resulting structural parameters in Table~\ref{tab:dwarfs}. The quoted $r_h$ is the best-fit elliptical half-light radius along the semi-major axis. Uncertainties are determined via bootstrap resamples. We check our results by repeating the calculations with the same set of stars, but with a limit one magnitude fainter. The derived structural parameters using both samples are consistent within the uncertainties.

Our results for Scl-MM-dw1 are in good agreement with those derived from Magellan$+$Megacam imaging \citep{Sand14}: we find $r_h=18.8\pm1.8$~arcsec with an ellipticity of $0.20\pm0.07$ while their value is $16.8\pm2.4$~arcsec with an ellipticity of $<$$0.42$. Compared to the structural parameters derived with integrated light from DES data \citep[$r_h=5.5\pm0.4$~arcsec, $\epsilon=0.39\pm0.04$,][]{Martinez-Delgado21}, our ML analysis for Scl-MM-dw3 suggests a similar size ($r_h=6.6\pm1.8$~arcsec) and a more elongated shape ($\epsilon=0.57\pm0.12$). Due to its large physical size ($r_h=3.2\pm0.5$~arcmin), the {\it HST} FoV is too small to derive robust structural parameters of Scl-MM-dw2. Therefore, we adopt the results from our Magellan$+$Megacam imaging \citep{Toloba16}, and do not attempt to revisit its structural parameters here. 

\subsection{Luminosity \label{sec:lum}}
We derive absolute magnitudes for our objects by using the same procedure as in \citet{MutluPakdil2018}, as was first described in \citet{Martin08}. First, we produce a well-populated CMD (of $\sim$20,000 stars) in {\it HST} filters, including our completeness and photometric uncertainties, by using an old, metal-poor PARSEC isochrone\footnote{We use an isochrone with age 12 Gyr and [M/H]$=-2.0$~dex. Using one with [M/H]$=-1.5$~dex or [M/H]$=-2.5$~dex gives a result consistent within the uncertainties.} and its associated luminosity function assuming a Kroupa IMF \citep{Kroupa2001}. We then randomly select the same number of stars from this artificial CMD as was found from our profile fits. We obtain the total luminosity by summing the flux of these stars, and extrapolating the flux of the faint, unresolved component of the galaxy from the adopted luminosity function. We calculate 1000 realizations in this way, and take the mean as our absolute magnitude and its standard deviation as the uncertainty. To account for the uncertainty on the number of stars (assuming Poisson statistics), we repeat this operation 100 times, varying the number of stars within its uncertainty, and use the offset from the best-fit value as the associated uncertainty. These error terms and the distance modulus uncertainty are then added in quadrature to produce our final uncertainty on the absolute magnitude. The final values can be found in Table~\ref{tab:dwarfs}.

Our {\it HST}-based luminosity measurement gives a fainter value for Scl-MM-dw1 ($M_V=-8.75\pm0.11$ versus the Magellan-based luminosity of $M_V$$=$$-10.3\pm0.6$, \citealt{Sand14}). It is worth noting that the Magellan-based luminosity was derived using a large aperture around the extent of the dwarf, and the existence of a bright star and several background galaxies in the vicinity of Scl-MM-dw1 (see Figure~\ref{fig:position}) appears to be the reason for a brighter result in the ground-based data. For Scl-MM-dw3, our luminosity measurement ($M_V=-7.24^{+0.26}_{-0.21}$) is consistent with the one reported in \citet[][$M_V=-7.04\pm0.2$]{Martinez-Delgado21}, within the uncertainties. For Scl-MM-dw2, similar to its structural parameters, we opt to use the Magellan-based luminosity \citep[$M_V$$=$$-12.1\pm0.5$,][]{Toloba16} due to its large physical size and the small ACS FoV. 

\subsection{\hi~Gas Limits}

\citet{Sand14} used data from the \hi~Parkes All-Sky Survey (HIPASS; \citealt{Barnes2001}) to constrain the \hi~content of Scl-MM-dw1, with no detection, and a 3-$\sigma$ \hi~gas mass upper limit of $log(M_{HI}/M_{\odot}) \lesssim 6.5$. Likewise, \citet{Toloba16} first checked the HIPASS spectra for Scl-MM-dw2, and found a tentative $\sim3.7\sigma$ \hi~emission peak along its line of sight. However, the authors later obtained much deeper \hi~ observations on the Green Bank Telescope, and did not detect any \hi~gas in emission, which constrains the gas mass of Scl-MM-dw2 to a 5-$\sigma$ upper limit of $log(M_{HI}/M_{\odot}) < 5.1$ \citep{Toloba16}.
 
Similarly, we investigate the possibility of \hi~gas associated with our three discoveries by using the HIPASS data, and find no evidence of a detection in any of them. The typical noise in these spectra is 18 mJy in 13.2 km~s$^{-1}$ channels. That gives a 3-$\sigma$ flux limit, assuming any real source would span 2 channels, of 1 Jy~km~s$^{-1}$. Converting this to an \hi~mass limit gives us a 3-$\sigma$ \hi~gas mass upper limit of $log(M_{HI}/M_{\odot}) \lesssim 6.5$ (see Table~\ref{tab:dwarfs}). This limit is consistent with the dwarfs being gas-poor, similar to other faint Local Volume dwarf satellites which reside within the virial radius of their primary galaxy \citep[e.g.,][]{Grcevich2009,Spekkens14,Karunakaran20}. However, given their low stellar masses, their gas content would not be detectable even if it was relatively high. 

\begin{figure*}
\centering
\includegraphics[width = \linewidth]{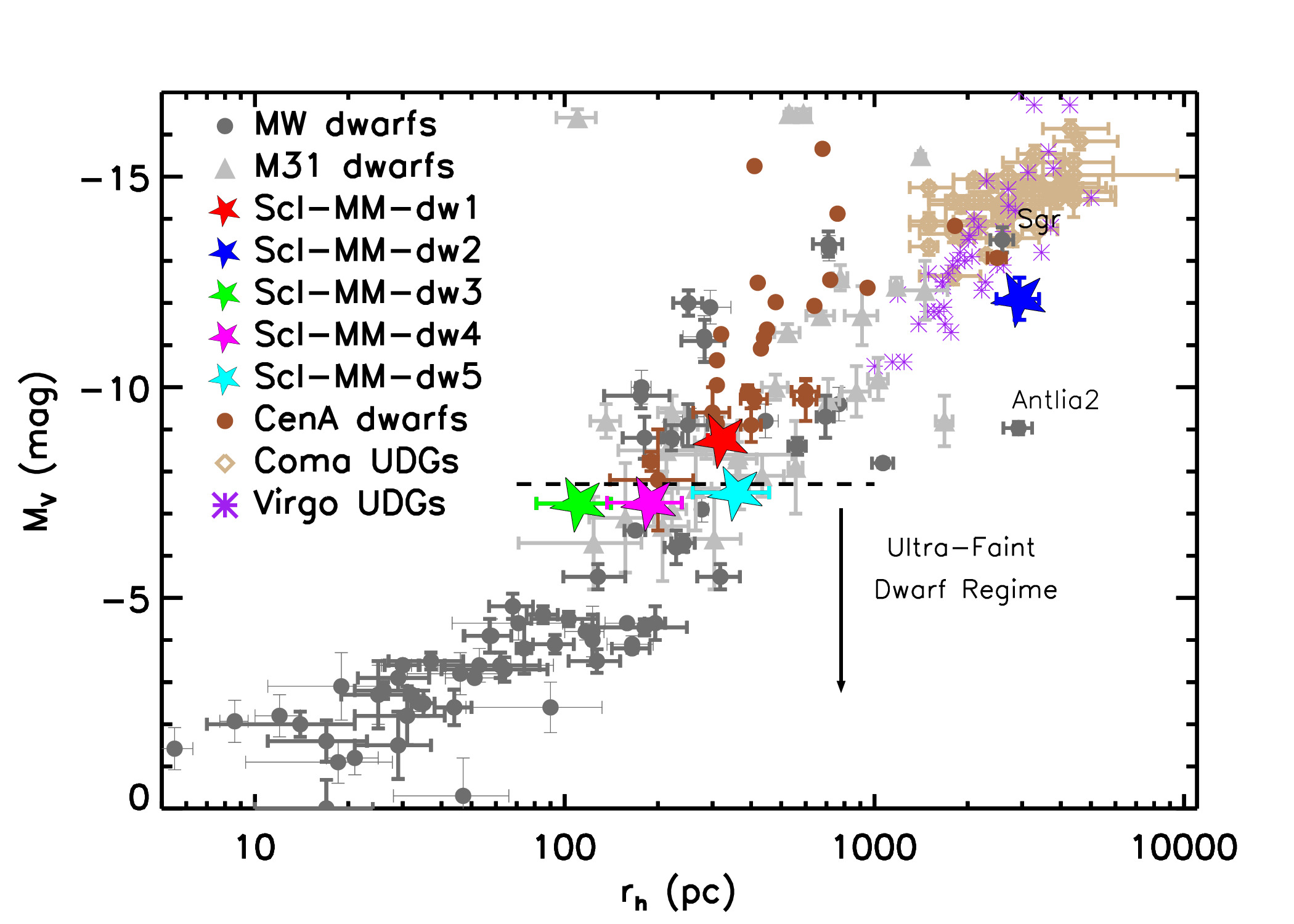}
\caption{Absolute V-band magnitude as a function of half-light radius for NGC~253 dwarfs, relative to MW/M31 dwarf galaxies, Cen~A dwarfs \citep{Sharina2008,Crnojevic19}, and the faint, diffuse galaxies found in Virgo and Coma by \citet{Lim2020} and \citet{vanDokkum2015}, respectively. The newly discovered NGC~253 dwarfs have similar properties to those of LG dwarfs. Scl-MM-dw2 is a slight outlier with a large half-light radius and low surface brightness for its absolute magnitude, similarly to Sagittarius in the Local Group and the recently discovered diffuse Virgo galaxies. The new NGC~253 satellites are among the faintest systems discovered beyond the Local Group. \label{fig:compare}}
\end{figure*}

\section{Discussion and Conclusions}\label{sec:conclude}

In this work, we report the discovery of three ultra-faint dwarf satellite galaxies of NGC~253 in a visual search of the Magellan/Megacam images taken as part of PISCeS, our panoramic imaging campaign to find faint substructure within $\lesssim$100~kpc of NGC~253. This brings the total number of NGC~253 satellites uncovered by PISCeS to five. We present {\it HST} follow-up of these five dwarfs, confirm their nature, and firmly establish their membership with NGC~253 by deriving TRGB distances.

We estimate the structural parameters and luminosities of NGC~253 dwarfs and compare them with those of the Local Group dwarfs as well as Cen~A dwarfs from our own PISCeS program, and with the ultra diffuse galaxies in the Virgo and Coma clusters (see Figure~\ref{fig:compare}). They are all comparable to MW and M31 dwarf galaxies. However, Scl-MM-dw2 is a slight outlier: it is one of the most extended and least dense objects known at its luminosity. It is similar to Sagittarius in the Local Group and the recently discovered diffuse Virgo galaxies.  It is likely undergoing a tidal interaction \citep[see discussion in][]{Toloba16}, as it has a high ellipticity ($\epsilon$$\approx$0.66) and is elongated in the direction of NGC~253. 

Our luminosity measurement places Scl-MM-dw1's luminosity near the faint end of those of the classical dSphs in the MW and M31. The MW satellite most similar to Scl-MM-dw1 is Draco ($M_V=-8.8\pm0.3$~mag; $r_h=221\pm26$~pc; $\epsilon=0.31\pm0.02$, \citealt{McConnachie2012}). Our latest three discoveries are well within the ultra-faint dwarf regime ($M_V\gtrsim-7.7$; \citealt{Simon2019}) in the size-luminosity plane. Their structural parameters and luminosities are comparable to MW ultra-faint dwarf Eridanus~II ($M_V=-7.1\pm0.3$~mag; $r_h= 277\pm14$~pc; $\epsilon=0.48\pm0.04$, \citealt{Crnojevic2016eri}). They are among the faintest dwarf satellites identified beyond the Local Group via a systematic search, demonstrating the effectiveness of PISCeS in extending the faint end of the satellite luminosity function for NGC~253.  

The five PISCeS dwarfs are all predominantly old and metal-poor stellar systems (age$\sim$12~Gyr; [M/H]$\lesssim$$-$1.5~dex). This is not surprising especially for our three ultra-faint discoveries: the ultra-faint dwarfs are uniformly old, with nearly all of their stars forming in the early universe, thus considered as pristine fossils from the era of reionization \citep[e.g.,][]{Bovill2011,Salvadori2009,Brook2014,Brown2014}. Also, dwarf satellites within $\sim100$~kpc of their primary galaxy are highly susceptible to loss of their cold gas through tidal or gas dynamical interactions \citep[e.g.,][]{Grebel2003,Mayer2006}. Our non-detection in HIPASS is consistent with this picture. 

Compared to our new discoveries, the stellar populations of Scl-MM-dw1 and Scl-MM-dw2 are relatively complex with some evidence of AGB stars. While Scl-MM-dw1 has a handful of AGB consistent with a population of $\gtrsim6$--$8$~Gyr, Scl-MM-dw2 shows a clear population of AGB stars which are $\sim$1--2~Gyr old with [M/H]$\sim$$-$1.0~dex. Unfortunately, the small {\it HST} FoV prevents us from exploring further any trends in the spatial distribution of RGB and AGB stars in Scl-MM-dw2.

\citet{Martinez-Delgado21} recently performed a visual NGC~253 satellite search using the DES data, and reported three new dwarf candidates, one of which is the same object we independently discovered and named as Scl-MM-dw3. The other two are located outside our PISCeS footprint, and their discoveries suggest that there might be more satellites to be discovered at larger radii from NGC~253. Moreover, the authors suggested the possible existence of a spatially flattened and velocity-correlated satellite galaxy system around NGC~253, which might point to an infalling filament or tidal origin. This flattened structure is $31\pm5$~kpc thick (i.e, thickness is here defined as the root-mean-square height from the best-fit plane) with the minor-to-major axis ratio of $0.14\pm0.03$, therefore comparable to the satellite planes found around the MW and M~31 \citep[e.g.,][]{Pawlowski2013,Ibata13}. Roughly $\sim$30\% of our survey footprint falls on this proposed linear structure, and 4 of our 5 dwarfs are consistent with lying along this plane.  Follow-up velocities of these dwarf galaxies \citep[e.g.][]{Toloba16_kinematics,Toloba16_kinematics2} will further elucidate the substructure properties of NGC253 and the Sculptor group.

We conclude by highlighting the crucial role played by PISCeS in identifying ultra-faint dwarf galaxies beyond the Local Group. A future paper will be dedicated to our overall satellite detection efficiency and the luminosity function of NGC~253.  This will provide a unique opportunity to study the  faint end of the satellite luminosity function in a new, more isolated environment than the Local Group. 

\begin{table*} 
\centering
\normalsize
\caption{{\it HST}-derived properties of NGC~253 dwarfs \label{tab:dwarfs}}
\begin{tabular}{lccccc}
\tablewidth{0pt}
\hline
\hline
Parameter & Scl-MM-dw1 & Scl-MM-dw2 & Scl-MM-dw3 & Scl-MM-dw4 & Scl-MM-dw5 \\
\hline
R.A. (deg) & 11.89643$\pm$2.0\arcsec & 12.57108$\pm$4.0\arcsec$^{*}$ & 11.77950$\pm$1.4\arcsec  & 13.45476$\pm$1.6\arcsec & 12.60776$\pm$1.2\arcsec \\
Dec. (deg) & $-$26.38971$\pm$2.0\arcsec & $-$24.74961$\pm$7.3\arcsec$^{*}$ & $-$23.95573$\pm$0.6\arcsec & $-$25.47442$\pm$1.8\arcsec & $-$26.72726$\pm$2.5\arcsec  \\
F814W$_{\rm TRGB}$ (mag) & 23.72$\pm0.33$ & 23.72$\pm0.03$ & 23.69$^{+0.06}_{-0.15}$ & 24.05$^{+0.06}_{-0.15}$ & 23.94$^{+0.08}_{-0.13}$ \\
$m-M$ (mag)           & $27.73\pm0.33$      &  $27.74\pm0.07$ &  $27.70^{+0.09}_{-0.18}$   &   $28.07^{+0.09}_{-0.18}$ & $27.95^{+0.10}_{-0.15}$    \\  
$D$ (Mpc)             & $3.53\pm0.55$      & $3.53\pm0.11$ &  $3.48^{+0.14}_{-0.28}$      &  $4.10^{+0.16}_{-0.32}$      & $3.90^{+0.18}_{-0.27}$       \\ 
$D_{proj}$ (kpc)      & 66        & 50    & 81        &  86       & 96           \\
$M_{V}$ (mag)         & $-8.75\pm0.11$   & $-12.10\pm0.50^{*}$ &  $-7.24^{+0.26}_{-0.21}$  &  $-7.26^{+0.27}_{-0.23}$ & $-7.50^{+0.28}_{-0.20}$ \\ 
$M_{\rm star} (M_\odot$) & $(4.3\pm0.5) \times10^5$   & $(0.9^{+0.6}_{-0.3}) \times10^7$ & $(1.1\pm0.2) \times10^5$ & $(1.1\pm0.3) \times10^5$ & $(1.4\pm0.3) \times10^5$\\
$log(M_{HI}/M_{\odot})$ &  $<6.5$    &  $<5.1$$^{*\dagger}$ &  $<6.5$  & $<6.5$ &    $<6.5$            \\
$r_{h}$ (arcsec)      &  18.8$\pm$1.8  & 194.4$\pm$30.6$^{*}$ &  6.6$\pm$1.8 &  9.5$\pm$2.6 & 19.0$\pm$5.2 \\     
$r_{h}$ (pc)   	      &  $321\pm31$ & $2940\pm460$$^{*}$ &  $111\pm30$  &  $188\pm51$  & $358\pm99$  \\  
$\epsilon$           &  $0.20\pm0.07$   & $0.66\pm0.06$$^{*}$ & $0.57\pm0.12$  &  $0.43\pm0.19$ & $0.66\pm0.11$ \\ 
Position Angle (deg)  &  $133\pm24$ & $31\pm3$$^{*}$ & $70\pm13$ & $80\pm46$ & $169\pm7$     \\
\hline
\end{tabular}
  \begin{tablenotes}
      \small
      \item  R.A.: the Right Ascension (J2000.0). DEC: the Declination (J2000.0). F814W$_{\rm TRGB}$: TRGB magnitude in F814W. $m-M$: the distance modulus. $D$: the distance of the galaxy in Mpc. $D_{proj}$: the projected distance to NGC~253 in kpc. $M_{V}$: the absolute V-band magnitude. $M_{\rm star}$: the stellar mass in solar mass, derived from the measured luminosity by assuming an average $V-$band mass-to-light of $M_{\text{star}}/L_{\text{V}} = 1.6$ \citep{Woo2008} appropriate for old stellar populations. $log(M_{HI}/M_{\odot})$: $3\sigma$ upper limits on the \hi~mass of each object. $r_{h}$: the elliptical half-light radius along the semi-major axis. $\epsilon$: ellipticity which is defined as $\epsilon=1-b/a$, where $b$ is the semiminor axis and $a$ is the semimajor axis.
      \item $^{*}$ The values are taken from \citet{Toloba16}.
      \item $\dagger$ This value was derived from deep \hi~observations from the Green Bank Telescope, and it corresponds to $5\sigma$ upper limits on the \hi~mass.
      
    \end{tablenotes}
\end{table*}

\acknowledgements

We are grateful to P. Bennet for useful discussions and tabulated data which were used in this paper. We also thank Annika Peter for useful comments.

Based on observations made with the NASA/ESA Hubble Space Telescope, obtained at the Space Telescope Science Institute, which is operated by the Association of Universities for Research in Astronomy, Inc., under NASA contract NAS5-26555. These observations are associated with program \# HST-GO-15938.  Support for program \# HST-GO-15938 and HST-GO-14259 was provided by NASA through a grant from the Space Telescope Science Institute, which is operated by the Association of Universities for Research in Astronomy, Inc., under NASA contract NAS5-26555. The Parkes telescope is part of the Australia Telescope which is funded by the Commonwealth of Australia for operation as a National Facility managed by CSIRO.

BMP is supported by an NSF Astronomy and Astrophysics Postdoctoral Fellowship under award AST-2001663.  DJS acknowledges support from NSF grants AST-1821967 and 1813708. Research by DC is supported by NSF grant AST-1814208. JDS acknowledges support from NSF grant AST-1412792. JS acknowledges support from NSF grant AST-1812856 and the Packard Foundation. KS acknowledges support from the Natural Sciences and Engineering Research Council of Canada (NSERC).

\vspace{5mm}
\facilities{Magellan:Clay (Megacam), HST (ACS), GALEX, Parkes}

\software{Astropy \citep{astropy13,astropy18}, The IDL Astronomy User's Library \citep{IDLforever}, DOLPHOT \citep{Dolphin2000}}


\end{document}